\documentclass[acmtog]{acmart}
\copyrightyear{2024}
\setcopyright{acmlicensed}
\acmConference[SA Conference Papers '24]{SIGGRAPH Asia 2024 Conference Papers}{December 3--6, 2024}{Tokyo, Japan}
\acmBooktitle{SIGGRAPH Asia 2024 Conference Papers (SA Conference Papers '24), December 3--6, 2024, Tokyo, Japan}
\acmDOI{10.1145/3680528.3687593}
\acmISBN{979-8-4007-1131-2/24/12}

\begin{document}
\def\sysName{\emph{DrawingSpinUp}}

\title{\sysName: 3D Animation from Single Character Drawings}

\author{Jie Zhou}
\orcid{0000-0002-3836-4163}
\email{jzhou67-c@my.cityu.edu.hk}
\affiliation{
    \institution{City University of Hong Kong}
    \country{China}
}

\author{Chufeng Xiao}
\orcid{0000-0001-6749-0161}
\email{chufengxiao@outlook.com}
\affiliation{
    \institution{Hong Kong University of Science and Technology}
    \country{China}
}

\author{Miu-Ling Lam}
\orcid{0000-0002-5333-7454}
\email{miu.lam@cityu.edu.hk}
\affiliation{
    \institution{City University of Hong Kong}
    \country{China}
}

\author{Hongbo Fu*}
\orcid{0000-0002-0284-726X}
\email{fuplus@gmail.com}
\affiliation{
    \institution{Hong Kong University of Science and Technology}
    \country{China}
}
\thanks{*Corresponding author.}

\renewcommand\shortauthors{Zhou et al.}

\begin{abstract}
Animating various character drawings is an engaging visual content creation task.
Given a single character drawing, existing animation methods are limited to flat 2D motions and thus lack 3D effects.
An alternative solution is to reconstruct a 3D model from a character drawing as a proxy and then retarget 3D motion data onto it. However, the existing image-to-3D methods could not work well for amateur character drawings in terms of appearance and geometry. 
We observe the contour lines, commonly existing in character drawings, would introduce significant ambiguity in texture synthesis due to their view-dependence. Additionally, thin regions represented by single-line contours are difficult to reconstruct (e.g., slim limbs of a stick figure) due to their delicate structures.
To address these issues, we propose a novel system, \sysName, to produce plausible 3D animations and breathe life into character drawings, allowing them to freely spin up, leap, and even perform a hip-hop dance. 
For appearance improvement, we adopt a removal-then-restoration strategy to first remove the view-dependent contour lines and then render them back after retargeting the reconstructed character.
For geometry refinement, we develop a skeleton-based thinning deformation algorithm to refine the slim structures represented by the single-line contours.
The experimental evaluations and a perceptual user study show that our proposed method outperforms the existing 2D and 3D animation methods and generates high-quality 3D animations from a single character drawing.
Please refer to our project page (\url{https://lordliang.github.io/DrawingSpinUp})
for the code and generated animations.

\end{abstract}

%
%

%
%


\begin{teaserfigure}
  \centering
  \includegraphics[width=0.96\textwidth]{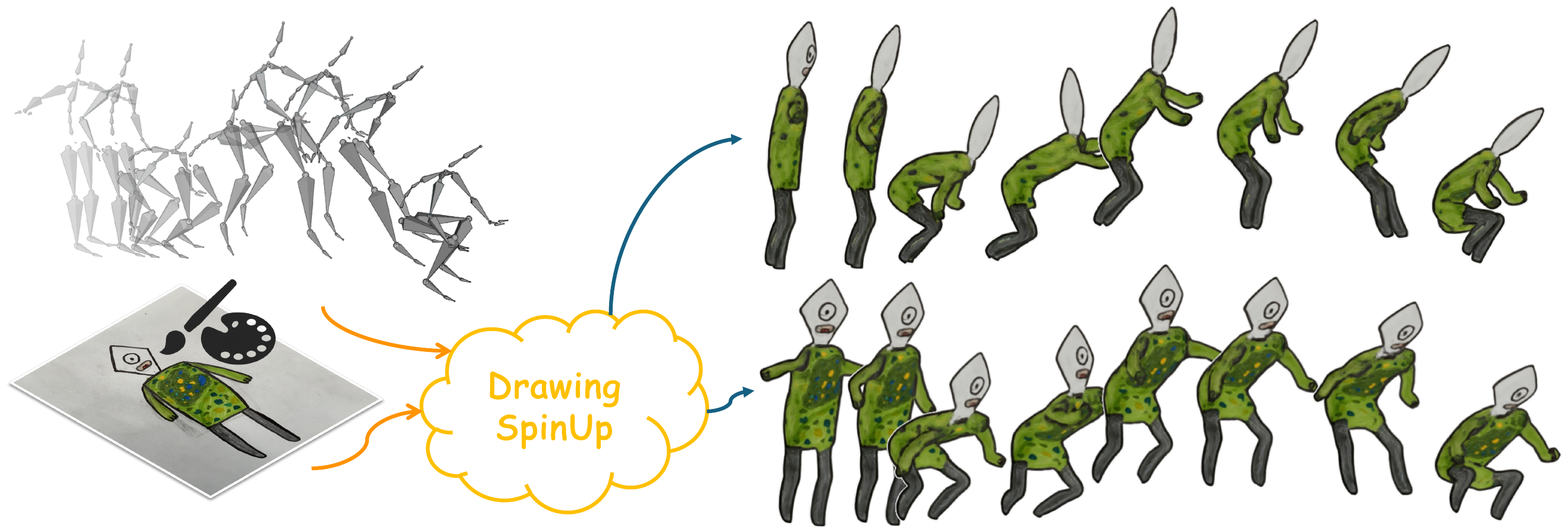}
  \vspace{-3mm}
  \caption{Our \sysName~produces visually vivid 3D character animations (Right) given single input drawings (Bottom-Left) and target motions (Top-Left).}
  \label{fig:teaser}
\end{teaserfigure}

\maketitle

\section{Introduction}
\label{sec:intro}
\begin{figure*}[htbp]
  \centering
  \includegraphics[width=0.96\textwidth]{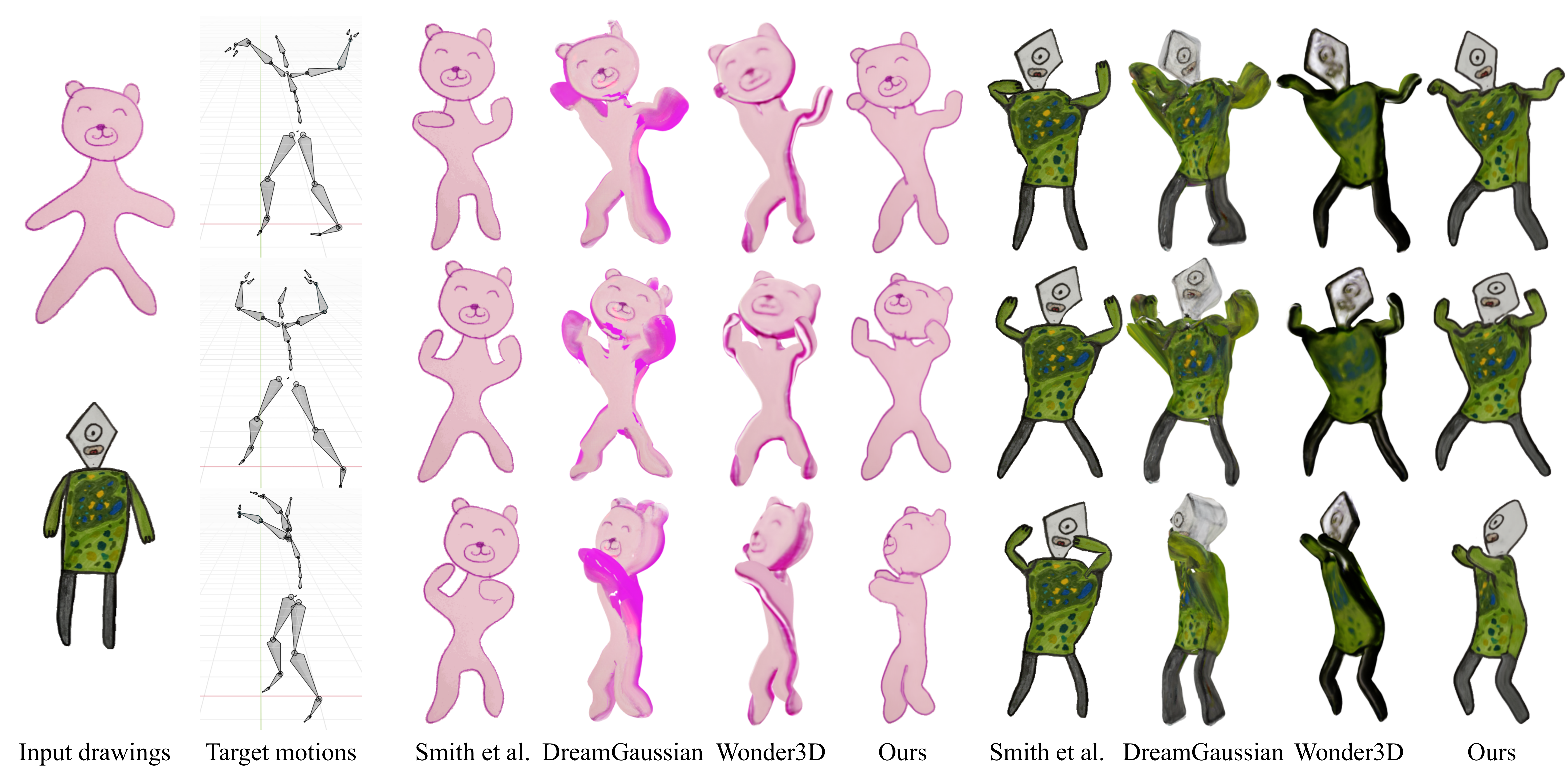}
  \vspace{-3mm}
  \caption{Our \sysName~produces visually more pleasing character animation results given input drawings and target motions than the existing 2D and 3D animation techniques.}
  \vspace{-3mm}
  \label{fig:existing}
\end{figure*}
Character drawing plays a crucial role in art and design, significantly enhancing storytelling, gaming, and animation~\cite{lasseter1987principles, librande1992example, huang2018character}. Beyond professionals, character drawing is a popular medium for amateurs, including children~\cite{cox2013children, smith2023method}, to unleash their inspiration by creating imagined characters like superheroes and fairy creatures. 
Given such a single drawing, could we bring the human-drawn character to life, e.g., making it run, leap, and dance on paper?

To animate a character drawing, existing methods (e.g.,~\cite{hornung2007character} and ~\cite{smith2023method}) typically retarget 2D motions onto the character by deforming its shape in 2D image space via as-rigid-as-possible (ARAP)~\cite{igarashi2005rigid}.
Unfortunately, these methods cannot work well for 3D motions beyond the 2D image plane, e.g., rotating the character's body or tilting its head, as shown in Fig.~\ref{fig:existing} (results of Smith et al.). This intuitive limitation stems from the lack of a 3D sense.
To afford 3D motions, a straightforward solution is to reconstruct 3D models as proxies from character drawings and then retarget 3D motions onto them.
However, existing image-to-3D methods (e.g., One-2-3-45~\cite{liu2024one}, One-2-3-45++~\cite{liu2024oneplus}, DreamGaussian~\cite{tang2023dreamgaussian}, Wonder3D~\cite{long2024wonder3d},  LRM~\cite{honglrm}) always struggle to reconstruct visually pleasing results from amateur character drawings due to the domain gap between photo-realistic images and human-drawn sketches, as shown in Fig.~\ref{fig:existing} (results of DreamGaussian and Wonder3D).

\begin{figure}[htbp]
  \centering
  \includegraphics[width=0.48\textwidth]{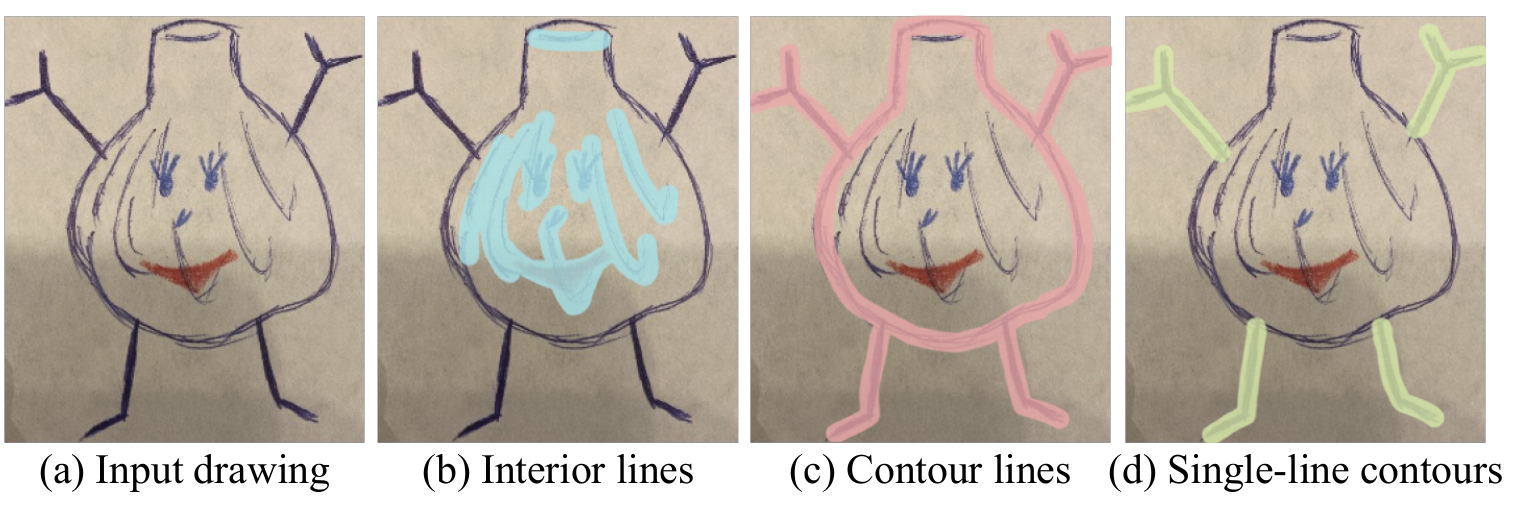}
  \vspace{-6mm}
  \caption{An example of {a character drawing with} diverse types of strokes, highlighted in blue (b), red (c) and green (d).}
  \vspace{-3mm}
  \label{fig:lines}
\end{figure}

Unlike photo-realistic images merely with texture details, we observe character drawings often exhibit diverse types of strokes, including interior lines and contour lines, as shown in Fig.~\ref{fig:lines} (b)-(c). Specifically, interior lines appear inside a character and represent texture patterns similar to those in photo-realistic images.
Different from interior lines, contour lines represent stylized character boundaries, which are view- and motion-dependent lines and are absent in photo-realistic images. 
Existing image-to-3D models are always trained on photo-realistic images, thus tending to mistake artistic contour lines for internal texture details. This ambiguity can lead to appearance degradation and even affect shape reconstruction quality, as shown in Fig.~\ref{fig:mesh_processing} (c). 
Moreover, these methods (e.g., Wonder3D~\cite{long2024wonder3d}) generally generate several fixed-view images and then fuse them to get a 3D model. Because this pipeline inherently handles contours as view-independent textures, re-training/fine-tuning them on 3D shapes with contour rendering still fails to address contour issues.
Additionally, single-line contours are commonly used to depict shapes and structures in an abstract manner, e.g., slim limbs in stick figures, as shown in Fig.~\ref{fig:lines} (d). Since such depiction is rarely seen in the training images, the models pre-trained on photo-realistic images might fail to reconstruct such delicate structures.

Based on the above observations, we present \sysName, the first 3D-aware animation system for generating non-planar animations from single character drawings given 3D motions. Our key idea is to recognize the contour lines and process them separately to adapt to the reconstruction prior of the pre-trained image-to-3D models.
Specifically, we adopt a removal-then-restoration strategy to handle the contour lines. We first design a network to remove contour lines in a character drawing and inpaint textures in the removed regions. Next, we utilize a pre-trained image-to-3D model to reconstruct a textured geometry as a proxy for 3D motion retargeting. Finally, we propose a geometry-aware stylization network to render view- and motion-dependent contour lines for each frame of the retargeted character and enhance the internal textures to match the style of the input image. In addition, to ensure the quality of geometry reconstruction, we develop a skeleton-based thinning deformation algorithm to refine the slim structures indicated by the single-line contours.
We select 120 representative amateur character drawing samples of various styles as test inputs from the Amateur Drawing Dataset collected by Smith et al.~\shortcite{smith2023method} to verify the efficiency of our method. The results of extensive experiments and a perceptual user study show that our \sysName~ can achieve plausible 3D reconstruction in geometry and appearance from single character drawings and generate vivid 3D animations, superior to existing 2D and 3D animation methods.

\section{Related Work}
In this section, we will summarize prior works closely related to character animation based on single drawings.

\subsection{2D Character Animation}
Igarashi et al.~\shortcite{igarashi2005rigid} introduce an as-rigid-as-possible (ARAP) deformation algorithm that enables users to intuitively deform a 2D shape by dragging keypoints, laying the foundation for animation in 2D image space.
Building on this, several subsequent methods~\cite{hornung2007character, smith2023method} typically project 3D motions onto a 2D plane to facilitate character deformation using ARAP.
Live Sketch~\cite{su2018live} adopts a stroke-preserving ARAP method for animating sketches while preserving the shape of user-specified strokes.
AniClipart~\cite{wu2024aniclipart} further develops a differentiable ARAP algorithm to optimize and warp a clipart to a new pose, guided by text-to-video priors. 
However, these 2D animation methods could not directly work for 3D animations due to single-view inputs and the lack of a 3D sense, as shown in Fig.~\ref{fig:existing} (results of Smith et al.).
Additionally, some data-driven methods approach this task as a conditional generation problem and utilize deep learning methods for character reanimation through motion transfer~\cite{chan2019everybody, hu2024animate, xu2024magicanimate}. However, these methods are still limited to generating results from a preset viewpoint.
In contrast, our proposed system can effectively transfer 3D motions to a character drawing by explicitly reconstructing a 3D model from the drawing and using it for motion retargeting, allowing viewpoint freedom.

\subsection{3D Character Animation}
To retarget 3D motions, many researchers have proposed to reconstruct 3D models as proxies from single-view drawings. 
Early methods often construct 3D meshes based on silhouette and skeleton via inflation~\cite{igarashi1999teddy, tai2004prototype, schmidt2005shapeshop, nealen2007fibermesh, buchanan2013automatic}.
Monster Mash~\cite{dvorovzvnak2020monster} incorporates 3D inflation with a layered deformation model to casually produce a smooth 3D mesh and animation given a single-view sketch.
Extended from Fibermesh~\cite{nealen2007fibermesh}, CreatureShop~\cite{zhang2022creatureshop} 
proposes an oblique-view modeling method to create fully-textured 3D character models by transferring textures between two intrinsically symmetric body parts. 
Parametric shape models have also been widely used in character reconstruction.
PhotoWakeUp~\cite{weng2019photo} proposes a 2D warping method to deform a skinned multi-person linear (SMPL) model~\cite{loper2015smpl} to fit the character silhouette of a single photo to create an animatable mesh.
ReenactArtFace~\cite{qu2023reenactartface} reconstructs a 3D artistic face through a 3D morphable model (3DMM)~\cite{paysan20093d} and a 2D parsing map from an input artistic image.
However, the above methods either ignore the back texture or simply mirror and duplicate the given front-view texture onto the back of a character, failing to address the texture issue effectively.
To address this issue, many data-driven methods~\cite{luo2023rabit, chen2023panic, peng2024charactergen} have been proposed to train generative models on a specific dataset to learn the missing textures. However, these methods are specifically tailored towards particular forms, such as formulated cartoon or anime characters. Unfortunately, there is a scarcity of large-scale pairs of hand-drawn character drawings and 3D assets for training since it is tedious and expensive to collect such data with subjective and artistic distortions.
Recent advances in novel view synthesis~\cite{liu2023zero, liu2024one, tang2023dreamgaussian, long2024wonder3d, honglrm} have brought a new solution, which is to exploit the powerful 3D prior capabilities of these pre-trained image-to-3D models to directly predict the missing textures.
In this paper, we utilize a pre-trained diffusion model~\cite{long2024wonder3d} to generate multi-view images to compensate for the lack of back textures.

\subsection{Contour Rendering}
As discussed in Section \ref{sec:intro}, non-photorealistic contour lines of character drawings may degrade reconstruction quality due to its nature of view- and motion-dependence. A few works have recognized this issue and managed to mitigate it. PAniC-3D~\cite{chen2023panic} proposes a line-infilling method to translate anime character images to render-like images more conducive to 3D reconstruction. It extracts lines with DoG operator, which deals with all lines equally, thus a facial landmark detector has to be used to keep certain lines around key facial features preserved.
Qu et al.~\shortcite{qu2023reenactartface} propose to synthesize contour lines for artistic faces by leveraging the input parsing map and a contour loss. Due to the use of uniform-thickness contour extraction strategy, their method might fail to handle the style of contour lines with inconsistent thickness and various textures.
For stylizing contours, existing stylization methods~\cite{benard2013stylizing, fivser2016stylit} that learn example-based styles are a potential solution, but they do not take the underlying 3D geometry into account and thus could not generate stylized contours with multi-view consistency.
Liu et al.~\shortcite{Liu_2021_ICCV} incorporate a 3D shape with a line drawing generated from Neural Contours~\cite{Liu_2020_CVPR} for stroke stylization. Instead of merely stylizing lines in \cite{Liu_2021_ICCV}, our task also needs to harmonize the generated contours with the interior textures of the given character drawing.
Inspired by the prior works, we first train a network to remove these view-dependent contour lines to prevent them from confusing the 3D reconstruction process and then design a geometry-aware stylization network to restore the contours, producing animations consistent with the input drawing style.

\section{Method}
\begin{figure*}[htbp]
  \centering
  \includegraphics[width=0.96\textwidth]{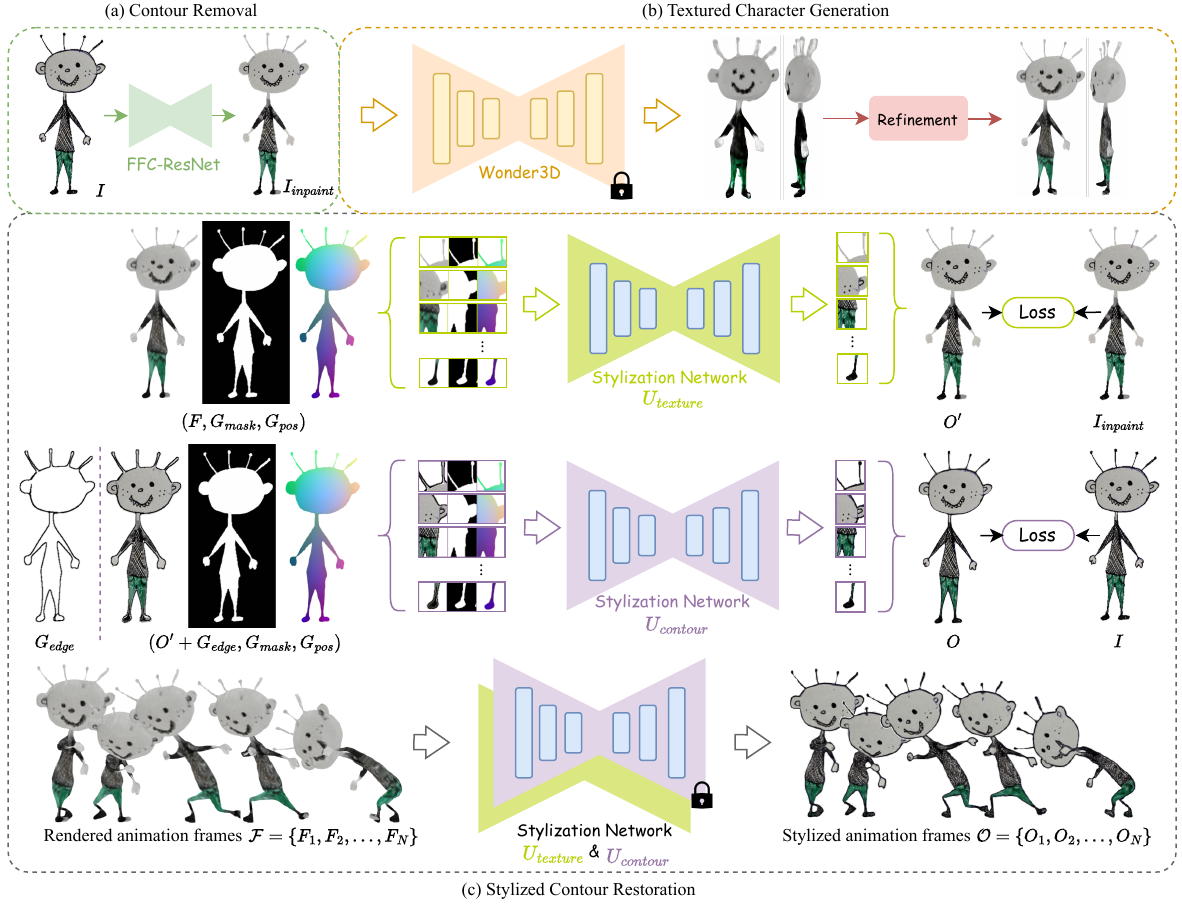}
  \vspace{-3mm}
  \caption{The pipeline of our \sysName. (a) We first remove and inpaint the contour region of the input drawing via an FFC-ResNet. (b) We use a pre-trained Wonder3D to generate a coarse 3D geometry and then refine its shape and texture. (c) We propose a two-stage geometry-aware stylization network to restore the original drawing style (including texture details and contour lines) for each animation frame.}
  \vspace{-3mm}
  \label{fig:pipeline}
\end{figure*}
We aim to build a system that can generate 3D animations by applying 3D target motions to a single character drawing. 
We follow the work of Smith et al.~\shortcite{smith2023method} to preprocess a human-drawn character drawing, including detection, segmentation, and pose estimation. Given a single character drawing, the foreground segmentation mask, the predicted joint keypoints, and a target 3D motion, our system \sysName~generates a vivid 3D animation while preserving the consistent artistic style with the input drawing.
Fig.~\ref{fig:pipeline} shows the pipeline of \sysName.
We first remove and inpaint the contour regions of the input drawing via an image-to-image translation network (Section \ref{sec:removal}). 
Next, we use a pre-trained diffusion model and a neural surface reconstructor to generate a coarse 3D geometry (Section \ref{sec:recon}).
Then we develop a shape refinement strategy to deal with the noisy surface and elongated structures depicted by thin strokes (Section \ref{sec:thin}).
We automatically rig the character based on the predicted joint keypoints and then retarget the given 3D motion onto it to generate an initial animation (Section \ref{sec:rig}).
Finally, we propose a geometry-aware stylization network to restore the original drawing style for each animation frame (Section \ref{sec:style}). We will present the details in the following subsections.

\subsection{Contour Removal}
\label{sec:removal}
The style and thickness of contours vary significantly across different drawings and can even be non-uniform within a single drawing. Therefore, using a distance transform with fixed parameters to extract contours is inadequate for all cases, leading to unclear results or excessive removal. To address this, we structure the contour prediction task as an image-to-image translation problem.
Given an input drawing $I$ and its segmented foreground mask $M$, we predict the corresponding contour mask $M_c$.
We use an FFC-ResNet~\cite{suvorov2022resolution} as the generator of our contour removal network. We choose FFC-ResNet because contour lines are typically found at the boundaries of objects, and Fast Fourier Convolution (FFC)~\cite{chi2020fast} has a large receptive field that covers the entire image, allowing 
for more accurate predictions of contour regions compared to vanilla convolution. 

We use the 3DBiCar dataset~\cite{luo2023rabit} as the training dataset. We render front-view images and contour lines of different thicknesses from textured 3D biped cartoon characters in 3DBiCar with Blender~\cite{blender}. Next, we stylize the contour lines with random colors and add them to the images to imitate the styles of amateur drawings.
\begin{figure}[htbp]
  \centering
  \includegraphics[width=0.48\textwidth]{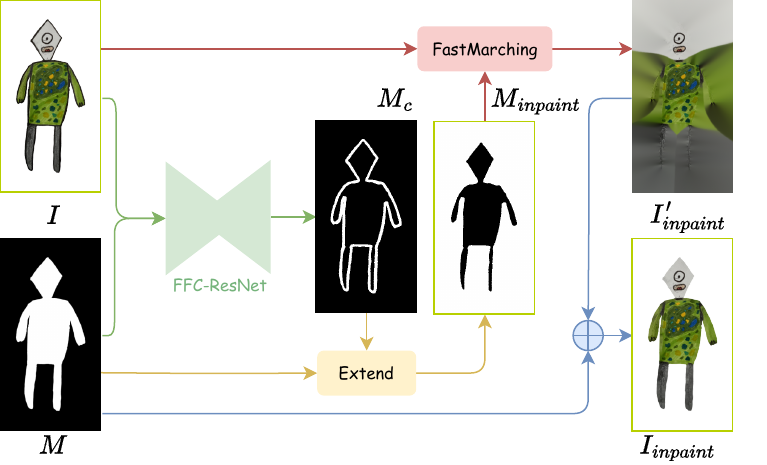}
  \vspace{-6mm}
  \caption{The process of contour removal.}
  \vspace{-3mm}
  \label{fig:contour_removal}
\end{figure}

After obtaining the predicted contour mask $M_c$, we remove the original contour by inpainting the masked region with the interior texture of the input drawing, as illustrated in Fig.~\ref{fig:contour_removal}. To eliminate the impact of the background color (always white) on the inpainting region, we extend $M_c$ by adding the background region $(1-M)$ into it to get the inpainting region mask $M_{inpaint}$:
\begin{equation}
M_{inpaint} = M_c \cup (1-M).
\end{equation}
Next, we inpaint the pixels within $M_{inpaint}$ heuristically based on a fast marching method~\cite{telea2004image}. That is, each pixel in $M_{inpaint}$ is replaced by a normalized weighted sum of the neighboring pixels in $(1-M_{inpaint})$. In a word, we compute the inpainted drawing $I_{inpaint}$ without contour by
\begin{equation}
\begin{aligned}
I_{inpaint}' &= FastMarching(I, M_{inpaint}); \\
I_{inpaint} &= I_{inpaint}' \cdot M + I \cdot (1-M).
\end{aligned}
\end{equation}
Please refer to supplemental materials for more details and a comparison of different methods for contour removal.

\subsection{3D Character Generation}
\subsubsection{Coarse Reconstruction}
\label{sec:recon}
\begin{figure}[htbp]
  \centering
  \includegraphics[width=0.48\textwidth]{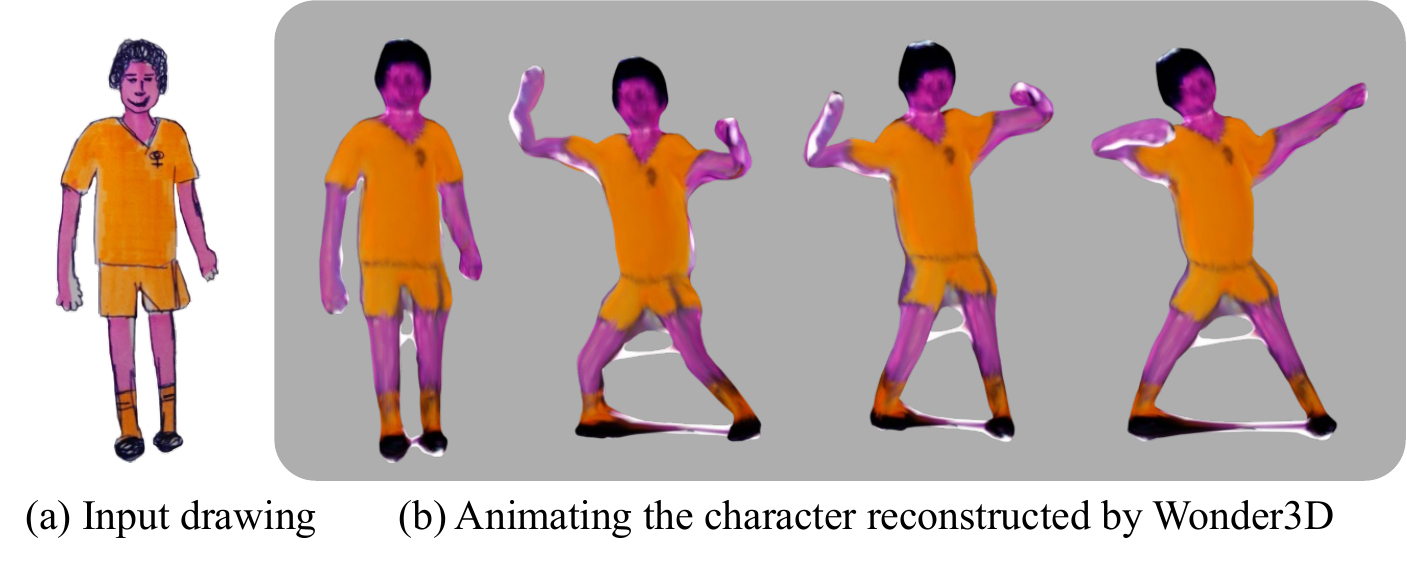}
  \vspace{-6mm}
  \caption{An example of surface adhesion.}
  \vspace{-6mm}
  \label{fig:adhesion}
\end{figure}

To reconstruct a 3D character from a single contour-free drawing, we use a pre-trained diffusion model, Wonder3D~\cite{long2024wonder3d}, to produce multi-view normal maps and color images. We choose Wonder3D as our backbone since it uses an orthographic camera setting, which helps to keep strong generalizations on amateur character drawings. We apply an off-the-shelf segmentation network, IS-Net~\cite{qin2022}, to segment these normal maps to get the corresponding foreground masks. Then we utilize a neural surface reconstructor, Instant-NSR~\cite{zhao2022human, instant-nsr-pl}, to reconstruct a textured geometry from these 2D representations. 
However, the obtained shape and texture are both far from satisfactory. As observed from Fig.~\ref{fig:mesh_processing} (d), the thin structures of the reconstructed shape might become much thicker than those in the input drawings. In some cases, there could even be adhesion on the surface, as shown in Fig.~\ref{fig:adhesion}, leading to geometry artifacts. Besides, the predicted texture appears blurry and loses details. These phenomena may result from the roughness of mask prediction and the misalignment of multi-view information. 

\subsubsection{Shape Cutting}
To refine the reconstructed shape, we employ front-view cutting on the queried SDF using the front-view mask $M$. Thus, the trimmed geometry can be defined by the $0$ level set:
\begin{equation}
\mathcal G = \left\{(x,y,z)\in \mathbb{R}^3 \mid f(x,y,z)\leq 0, M(X,Y)=1 \right\},
\end{equation}
where $f(\cdot)$ is the SDF and $(X,Y)$ is the projected 2D coordinates on the front-view plane of any 3D sampling point $(x,y,z)$.
Then we pass the level set to the Marching Cubes algorithm~\cite{lorensen1998marching} to extract the trimmed geometry, as shown in Fig.~\ref{fig:mesh_processing} (e). However, the front-view cutting operation can only change the silhouette of the front but not the thickness of the side.

\subsubsection{Skeleton-based Thinning Deformation}
\label{sec:thin}

To reduce the thickness of the side, we design a skeleton-based shape deformation algorithm to thin these regions without changing the front-view boundary. We solve it as a bi-harmonic problem~\cite{botsch2004intuitive}.
Given a trimmed geometry $\mathcal G$, we denote its vertices as $\mathbf{v}$ and its faces as $\mathbf{f}$. Thus, we can easily thin $\mathcal G$ by deforming it following
\begin{equation}
\begin{aligned}
\mathbf{v'}&=\mathbf{v}+\mathbf{d}; \\
\mathbf{d} &= \mathcal B(\mathbf{v}, \mathbf{f}, \mathbf{h}, \mathbf{d_h}).
\end{aligned}
\label{eq:bi-harmonic}
\end{equation}
where $\mathbf{d}$ represents a deformation field obtained from $\mathcal B(\cdot)$ that computes bi-harmonic maps using a uniform Laplacian operator~\cite{libigl}. We estimate the whole deformation field $d$ based on the known displacements $\mathbf{d_h}$, where $\mathbf{h}$ denotes handle indices.

\begin{figure}[htbp]
  \centering
  \includegraphics[width=0.4\textwidth]{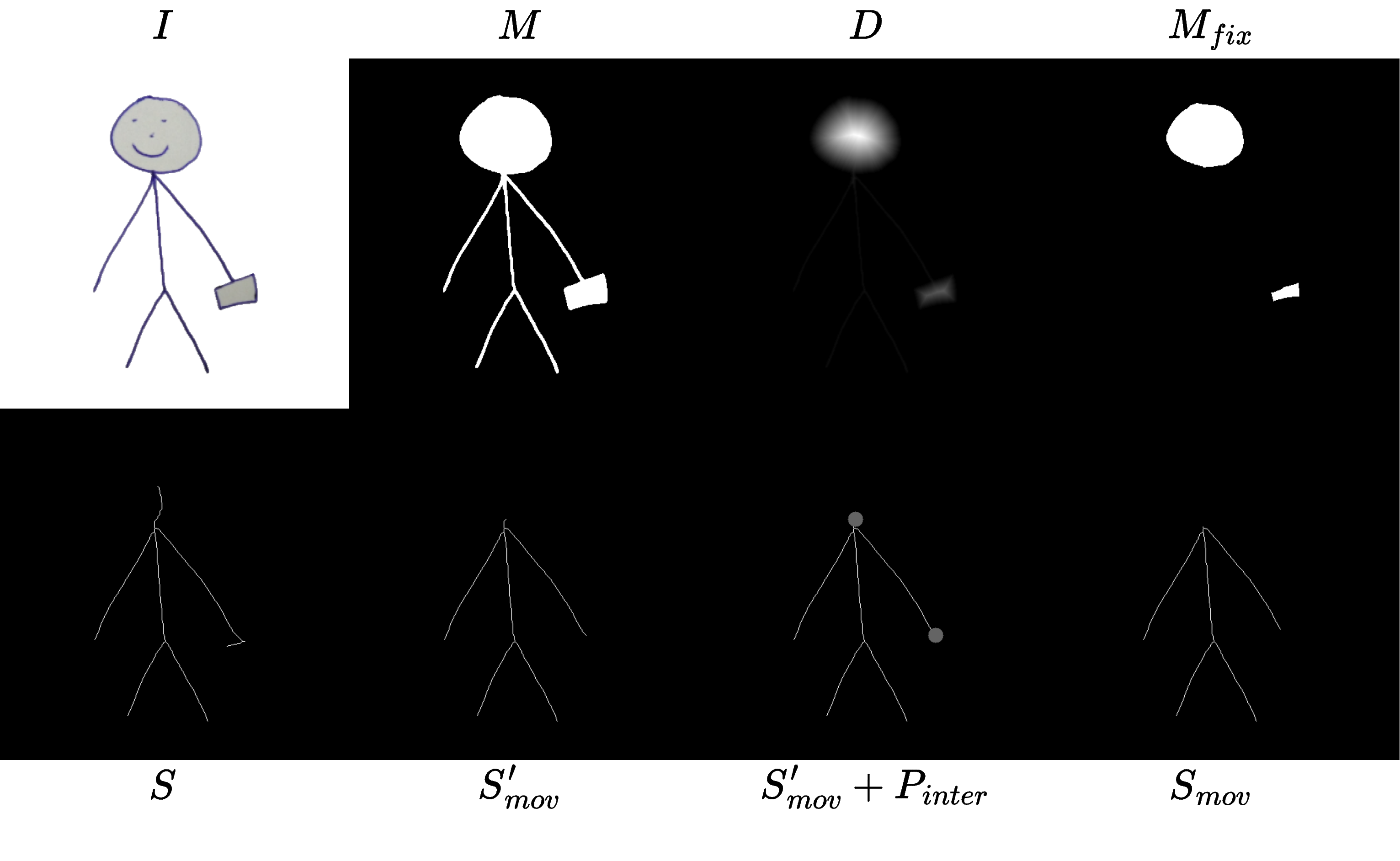}
  \vspace{-3mm}
  \caption{An example of handling vertex locations.}
  \vspace{-3mm}
  \label{fig:thin}
\end{figure}
\begin{figure*}[htbp]
  \centering
  \vspace{-3mm}
  \includegraphics[width=0.96\textwidth]{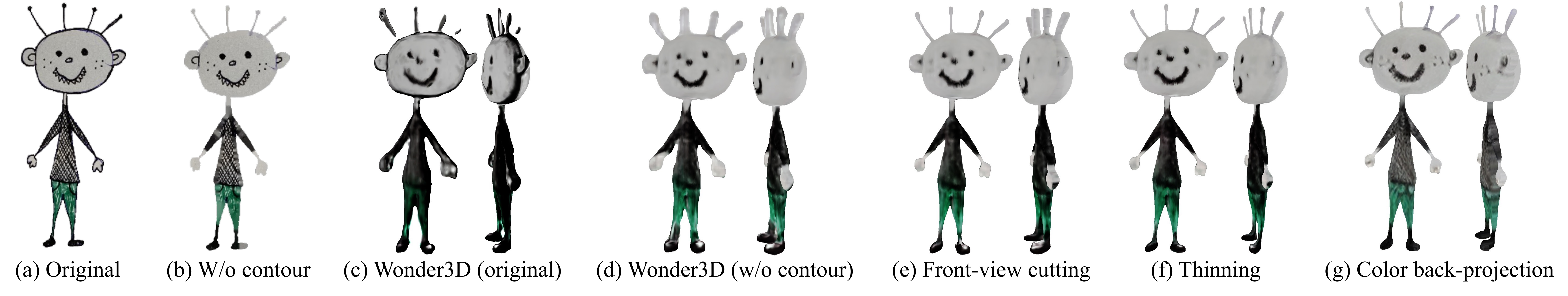}
  \vspace{-3mm}
  \caption{Results of different processing stages.}
  \vspace{-3mm}
  \label{fig:mesh_processing}
\end{figure*}
We next explain how we locate the handle vertices. We divide handle vertices into two categories (fixed and move-needed) to determine whether a local structure requires thinning. To distinguish them, as illustrated in Fig.~\ref{fig:thin}, we extract a distance map $D$ and a skeleton $S$ from $M$ via a medial-axis extraction algorithm \cite{lee1994building}.
Based on them, we can filter two masks $M_{fix}$ and $S_{mov}$ respectively, indicating the regions of fixing and move-needed vertices, following
\begin{equation}
\begin{aligned}
M_{fix} &= M \cdot (D \geq \theta_1), \\
S_{mov}' &= S \cdot (D \leq \theta_2), \\
S_{mov} &= IR(S_{mov}'),
\end{aligned}
\end{equation}
where $\theta_1$ and $\theta_2$ are distance thresholds and should meet $\theta_1 > \theta_2$. We use $\theta_1=11$ and $\theta_2=6$ in our experiments. Considering that the thinning deformation may affect adjacent areas (e.g., thinning the top of the neck may damage the face that is fixed), we remove the pixels within an intersection area $P_{inter}$ from $S_{mov}'$ via $IR(\cdot)$. Then we can easily distinguish fixed vertices $P_{fix}$ and move-needed vertices $P_{mov}$ based on $M_{fix}$ and $S_{mov}$. We compute the desired displacement for each vertex in $P_{mov}$ by querying the distance value from $D$. Finally, we update $P_{mov}$ through bilateral deformation following Eq. \ref{eq:bi-harmonic}. For sharp edges generated by cutting and thinning, we use Laplacian smoothing~\cite{vollmer1999improved} to smooth the surface.
Fig.~\ref{fig:mesh_processing} (f) shows an example of thinning the character's hair and limbs based on the proposed method.  Please refer to supplemental materials for more details. 

\subsubsection{Color Back-projection}
To improve the texture quality, we recolor each vertex by back-projecting multi-view color images onto 3D space, inspired by Peng et al.~\shortcite{peng2024charactergen}. In this paper, we focus exclusively on characters in a forward stance, a common posture in amateur drawings resembling an A/T-pose. This allows us to cover most textures by querying front and back view color images. For areas not visible from the front and back views, such as the inner side of the arms close to the body or the inner surface between the legs, we employ weighted colors from neighboring vertices for inpainting. Ultimately, we color each vertex to create a textured geometry, as illustrated in Fig.~\ref{fig:mesh_processing} (g).

\subsection{Rigging and Retargeting}
\label{sec:rig}
Given the contour-free 3D characters, we employ Mixamo's online rigging tool~\cite{mixamo} to automatically rig them to be animation-ready 3D assets. Mixamo uses a keypoint-based auto-rigging algorithm based on eight 2D joint keypoints on the front view. We directly reuse the joint keypoints offered by Smith et al.~\shortcite{smith2023method}. Given any humanoid 3D motion data, we can retarget it onto the rigged characters with Rokoko~\shortcite{rokoko-studio-live-blender} for animation rendering. As for skinning weight, we use Blender~\cite{blender} to automatically calculate the distance between each vertex and the closest bone and assign weights accordingly.

\subsection{Stylized Contour Restoration}
\label{sec:style}
Given the animated contour-free character, we can render a sequence of color frames $\mathcal{F}=\{F_1,F_2,...,F_N\}$ ($N$ is the sequence length).
As illustrated in Fig.~\ref{fig:pipeline} (c), we now restore the original drawing style (including texture details and contour lines) for each frame of $\mathcal{F}$, taking the stylized keyframe $I$ as a condition, to obtain the stylized frames $\mathcal{O}=\{O_1,O_2,...,O_N\}$.
Thus, we propose a two-stage geometry-aware stylization network to address this image-to-image translation task.

\subsubsection{Network Architecture}
\begin{figure}[htbp]
  \centering
  \vspace{-3mm}
  \includegraphics[width=0.48\textwidth]{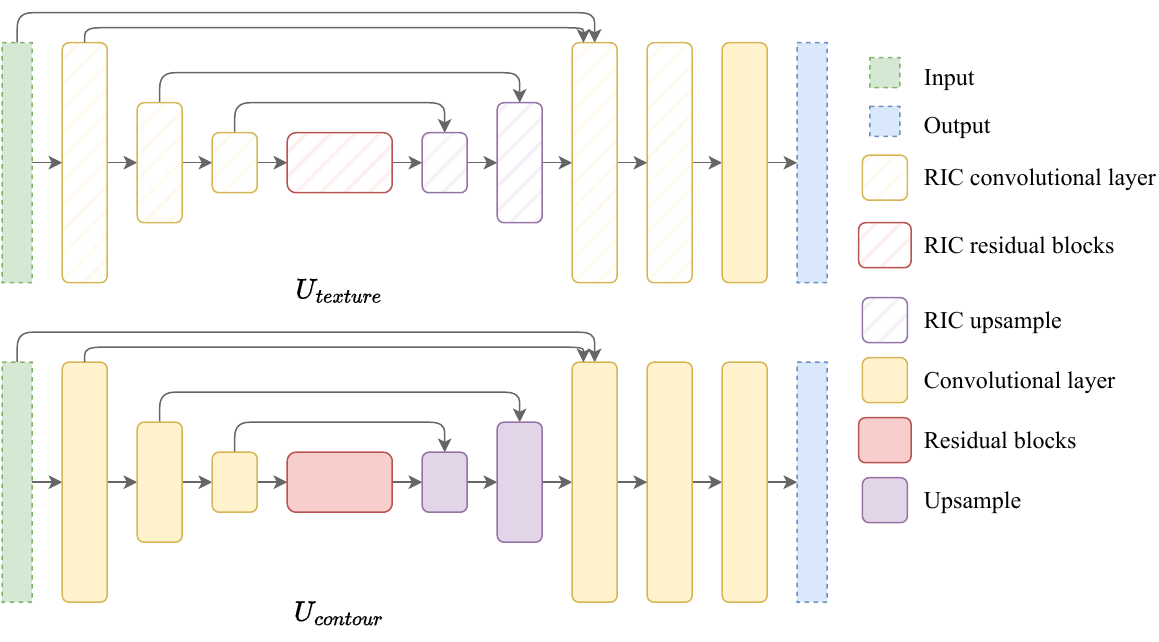}
  \vspace{-6mm}
  \caption{The network architectures of two U-Nets in our stylization network.}
  \vspace{-3mm}
  \label{fig:architecture}
\end{figure}

As illustrated in Fig.~\ref{fig:pipeline} (c), our stylization network is composed of two cascaded modified U-Nets of Futschik et al.~\shortcite{futschik2019real}. The first one $U_{texture}$ is responsible for restoring internal texture details, while the second one $U_{contour}$ focuses on restoring external contour lines. 
The network architectures of these two U-Nets are illustrated in Fig.~\ref{fig:architecture}.
Considering that many energetic motions may require the character to tilt or even head down while the vanilla convolutional layers are sensitive to rotation, as claimed by \cite{finnveden2021understanding, hao2022gradient, mo2024ric}, we replace all convolutional layers (except for the final layer) of $U_{texture}$ with rotation-invariant coordinate (RIC) convolutional layers \cite{mo2024ric} to enhance the stability of texture details.

\subsubsection{Geometry-aware Inputs}
Inspire by Jamri{\v{s}}ka et al.~\shortcite{jamrivska2019stylizing}, to maintain multi-view consistency, our stylization network takes as inputs four types of guidance channels, i.e., original color frame $F$, foreground mask $G_{mask}$, positional hint $G_{pos}$, and edge map $G_{edge}$. 
Specifically, $G_{pos}$ is obtained from the normalized $(x, y)$ coordinates of the character in a rest posture, providing the view-independent information, encouraging the source patches from $I$ to be transferred to similar relative positions in a color frame $F$. $G_{edge}$ is extracted from the Z-depth of each frame with the Canny edge detector~\cite{canny1986computational} and is used in the contour restoration stage to compensate for the view-dependent information.
More specifically, the first U-Net $U_{texture}$ takes $(F, G_{mask}, G_{pos})$ as input and generates the middle stylized frame $O'$. Then we overlap $G_{edge}$ on $O'$ and the second U-Net $U_{contour}$ takes $(O'+G_{edge}, G_{mask}, G_{pos})$ as input and generate the final stylized frame $O$.

\subsubsection{Patch-based Training}
Learning common knowledge via a generalized stylization network for contour restoration is difficult. This is because the artistic styles of character drawings have large variances in color, thickness, and stroke style. Thus, we adopt a patch-based training strategy based on limited training samples, following Texler et al.~\shortcite{texler2020interactive}.
We use the same training strategy for $U_{texture}$ and $U_{contour}$. Specifically, we randomly sample small $k\times k (where k=32)$ patches from all guidance channels and the ground truth. These patch pairs are then used to train the networks to generate corresponding patches with texture details or stylized contours.
We adopt a combination of L1 loss, adversarial loss, and VGG loss for supervised learning. The loss between $O'$ and $I_{inpaint}$ is used to optimize $U_{texture}$, while the loss between $O$ and $I$ is used to optimize $U_{contour}$. When inference, thanks to the settings of fully convolutional layers, we can feed each frame with full size to the network to finally restore internal texture details and stylized contours for a generated animation. 

\section{Experiments}
\subsection{Runtime}
Here we give the runtime of each stage on a single RTX 4090 GPU, including (1) contour removal (0.1s), 
(2) 3D character generation (2-3min), 
(3) online rigging (1-2min), 
(4) stylization network training (5-10min), 
(5) frame rendering in Blender with Eevee (0.1s/frame), and 
(6) stylization network inference (0.2s/frame).

The total training time is 10-15min for each character, and the inference time is 0.3s/frame. Note that we need to run Steps (1)-(4) only once to model the domain of a new character. Once the stylized network is trained, we can repeat Steps (5)-(6) to generate diverse animations given different 3D motions for the same character.

\subsection{Comparison to State-of-the-Art Methods}
\begin{figure}[htbp]
  \centering
  \includegraphics[width=0.48\textwidth]{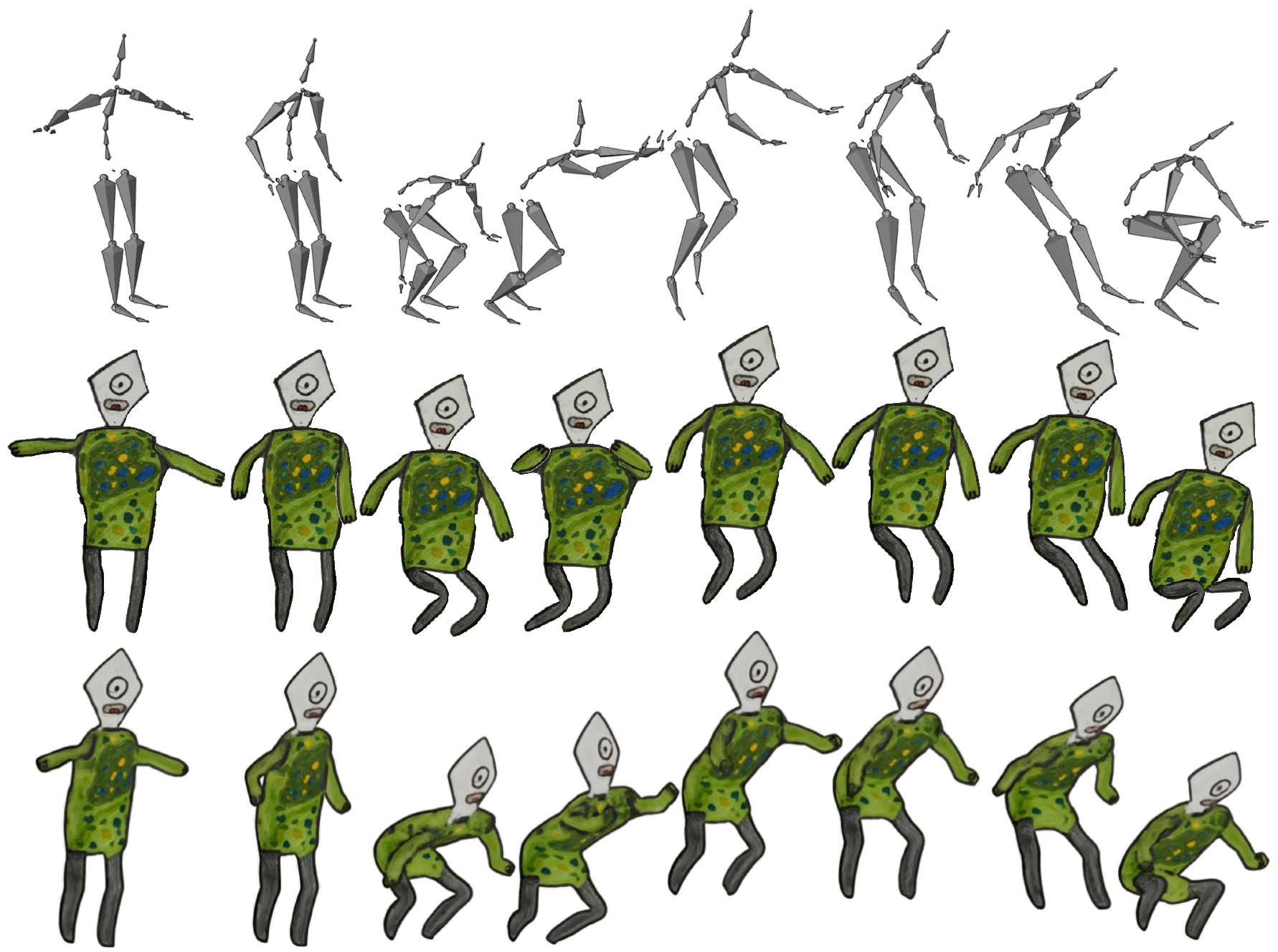}
  \vspace{-6mm}
  \caption{\sysName~ VS. Smith et al.~\shortcite{smith2023method}. From top to bottom: target motions (front-left view), Smith et al.'s result, and ours
  (front-left view).}
  \vspace{-3mm}
  \label{fig:comparison_2d}
\end{figure}
\paragraph{\sysName~ VS. Smith et al.~\shortcite{smith2023method}.}
Fig.~\ref{fig:comparison_2d} shows a comparison between Smith et al.'s method and ours, given the same 3D motion. We can observe the results generated by Smith et al. only exhibit planar motion {projected from the 3D motion} of the limbs, failing to capture the tilting of the body or the head of the character. Instead, our method can produce plausible 3D-aware animation that is faithful to the given 3D motion.

\begin{figure*}[hbtp]
  \centering
  \includegraphics[width=0.96\textwidth]{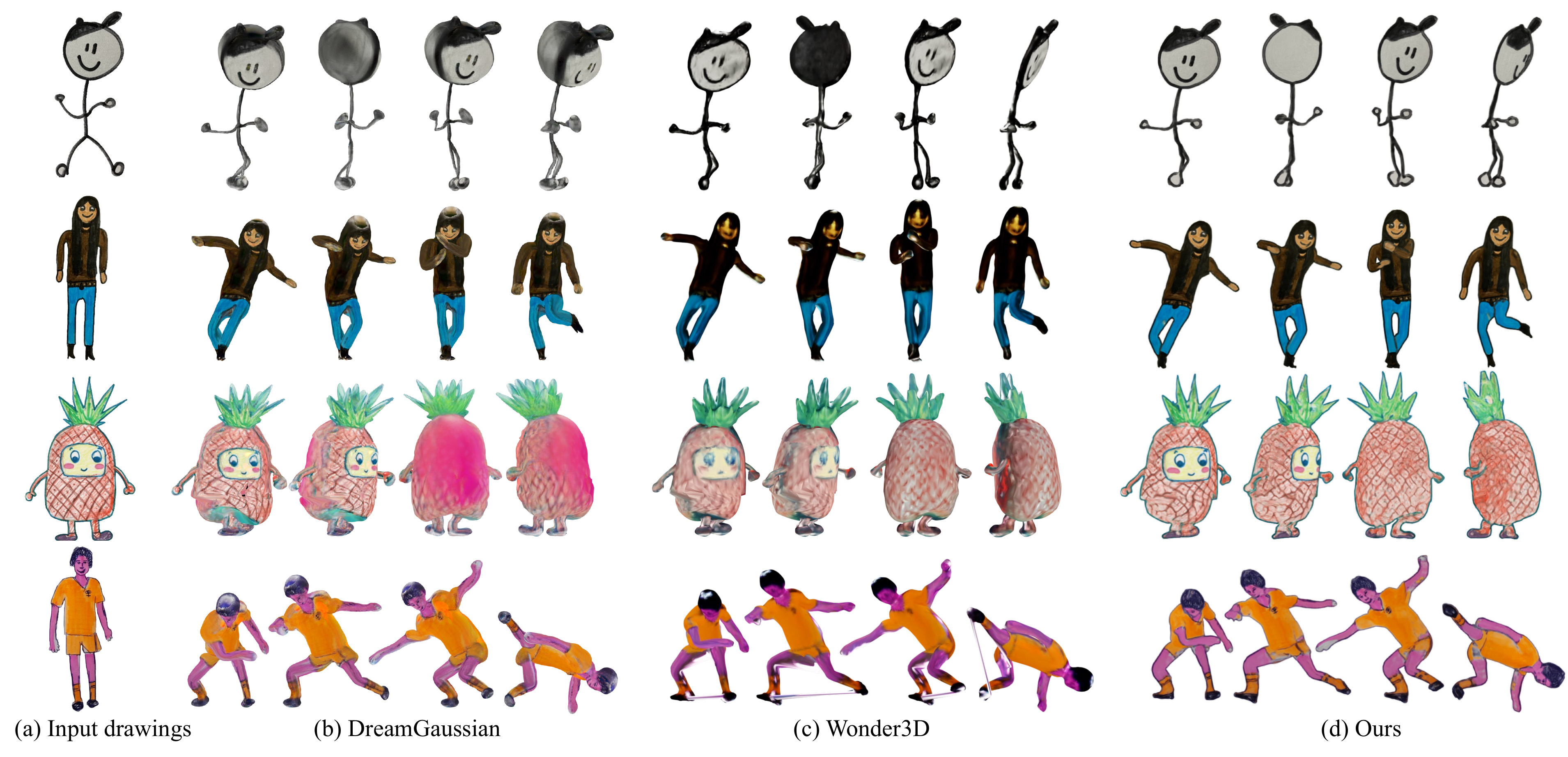}
  \vspace{-3mm}
  \caption{\sysName~ VS. other 3D-based animation methods.}
  \label{fig:comparison_3d}
\end{figure*}
\paragraph{\sysName~VS. Other 3D-based Animation Methods.}
To our knowledge, our work is the first to support 3D animation of amateur character drawings. Thus, we mainly compare our method with those based on different image-to-3D reconstruction backbones, including DreamGaussian~\cite{tang2023dreamgaussian} and Wonder3D~\cite{long2024wonder3d}. Fig.~\ref{fig:comparison_3d} shows the animated results based on their reconstructed characters {from the input drawings}. DreamGaussian tends to produce pure color but not details for novel-view textures, leading to serious artifacts at the back of characters. Wonder3D often generates messy textures due to the ambiguity caused by contours. Thanks to our contour removal network and stylization network, our method can maintain consistent styles with the input drawings and produce natural animations.

\subsection{Perceptual User Study}
We conducted a user study to evaluate the perceptual quality of our \sysName~ over the other compared methods via the following two metrics:
\begin{itemize}
\item Motion Consistency (MC): the alignment between the target motion and the generated motion; 
\item Style Preservation (SP): the preservation of the original style.
\end{itemize}
We conducted the user study via an online questionnaire with 15 groups of results by the compared methods (presented in a random order) and invited 53 human viewers to rate them in terms of the two metrics using a 5-point scale (1: "poor quality"; 2: "fair quality"; 3: "average quality"; 4: "good quality"; 5: "excellent quality"). The participants included a diverse demographic aged 23 to 55, with balanced gender representation. We recruited them via social media, ensuring a mix of expertise levels from beginners to professionals in animation and illustration.  
The average ratings for MC are 3.85 (Smith), 4.30 (DreamGaussian), 4.30 (Wonder3D), and 4.55 (Ours), respectively. The average ratings for SP are 4.18 (Smith), 3.65 (DreamGaussian), 3.50 (Wonder3D), and 4.53 (Ours), respectively.

We also conducted one-way ANOVA tests on the rating results and found a significant difference among these four methods for motion consistency (F=28.84, p<0.001) and style consistency (F=71.83, p<0.001). The further paired T-tests (with p<0.001) show that our method got a significantly higher rating in motion consistency than the other methods, i.e., Smith et al. (t=9.90), DreamGaussian (t=9.96) and Wonder3D (t=10.18). In terms of style consistency, our method also outperforms the other 3D-based methods, i.e., Smith et al. (t=4.31), DreamGaussian (t=20.74), and Wonder3D (t=23.954). The results clearly show that \sysName~ significantly outperforms the other three methods in terms of motion consistency and style preservation.

\subsection{Ablation Study}
\begin{figure}[htbp]
  \centering
  \includegraphics[width=0.48\textwidth]{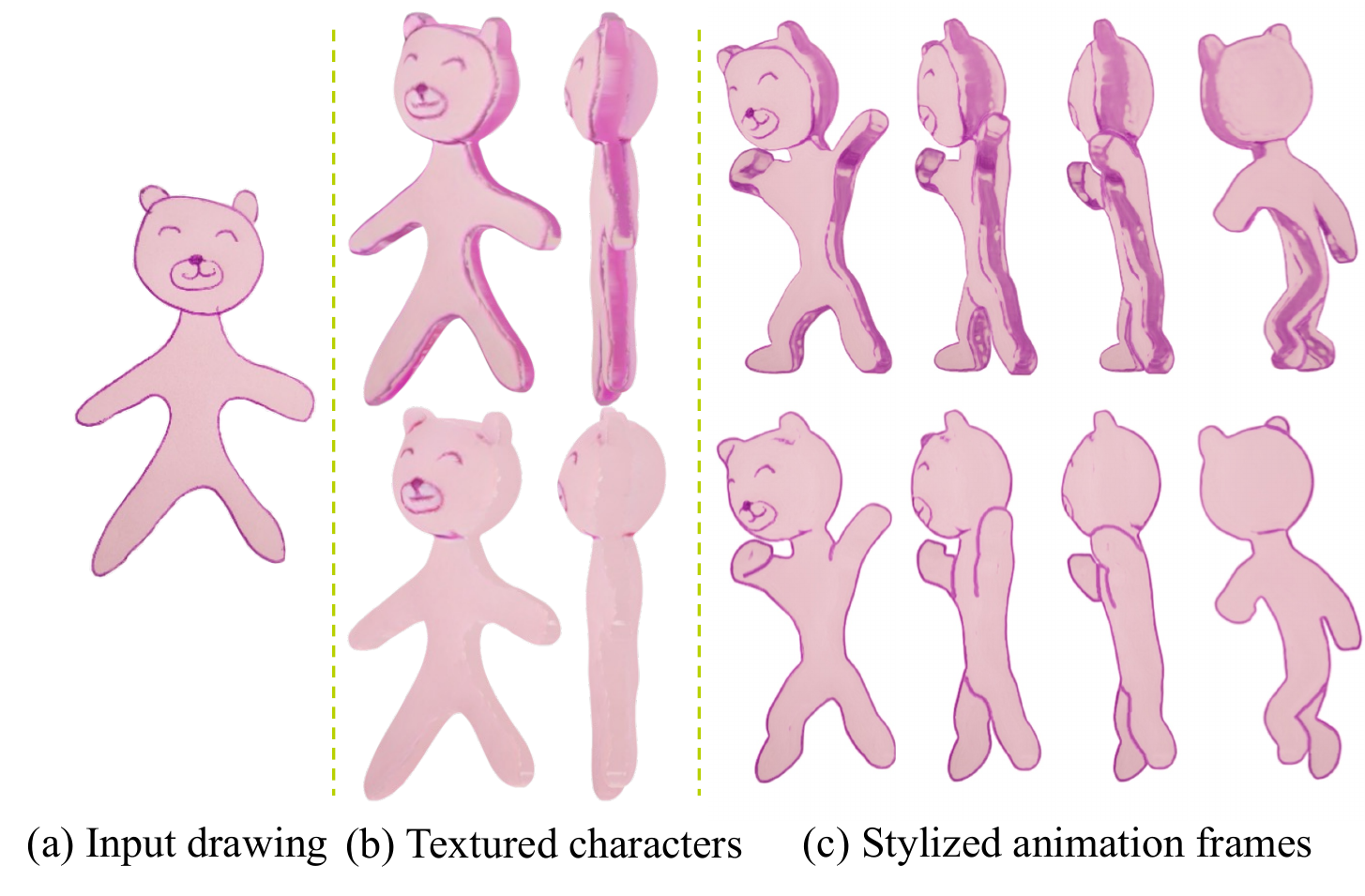}
  \vspace{-6mm}
  \caption{Comparison between w/o (Top) and w/ (Bottom) contour removal.}
  \vspace{-3mm}
  \label{fig:ab1}
\end{figure}
\paragraph{Contour Removal}
Fig.~\ref{fig:ab1} shows the impact of contour removal.
It can be seen that without this step, the contours would be considered as part of the internal texture, which cannot be solved by the following stylization step. In contrast, after contour removal, our method renders view-dependent contours to form a more plausible animation. 

\paragraph{Cutting and Thinning}
\begin{figure}[htbp]
  \centering
  \includegraphics[width=0.48\textwidth]{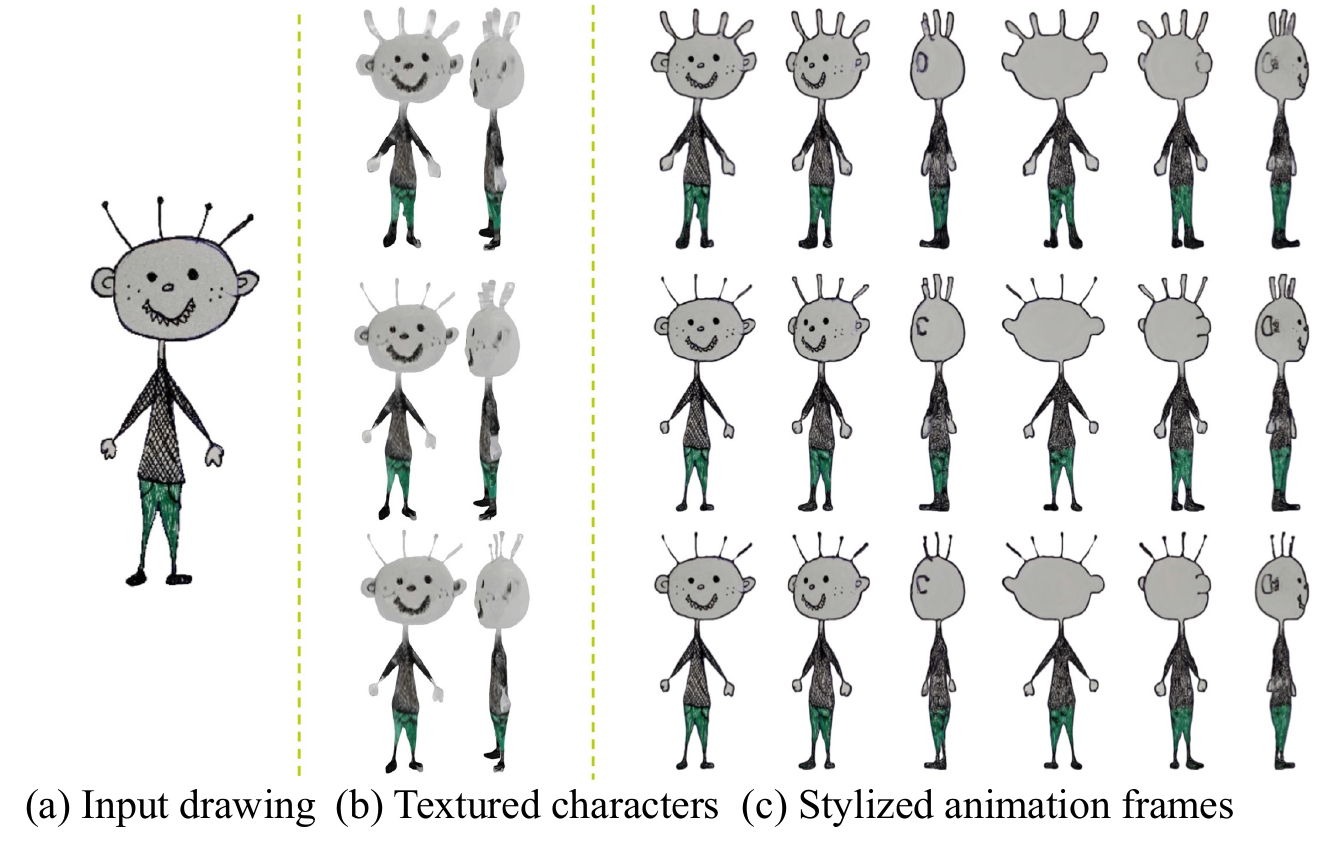}
  \vspace{-6mm}
  \caption{Comparison of three scenarios: no-cut-no-thin {(Top)}, only-cut {(Middle)}, and cut-and-thin {(Bottom)}.}
  \vspace{-3mm}
  \label{fig:ab2}
\end{figure}
We also present a comparison of three scenarios: no-cut-no-thin (Top), only-cut (Middle), and cut-and-thin (Bottom). As shown in Fig.~\ref{fig:ab2}, with no shape refinement or only cutting leads to inflated results for fine structures such as hair and limbs. In contrast, our cutting-and-thinning method can handle well the elongated structures depicted by slim strokes.

\paragraph{Rotation Invariance.}
\begin{figure*}[htbp]
  \centering
  \includegraphics[width=0.96\textwidth]{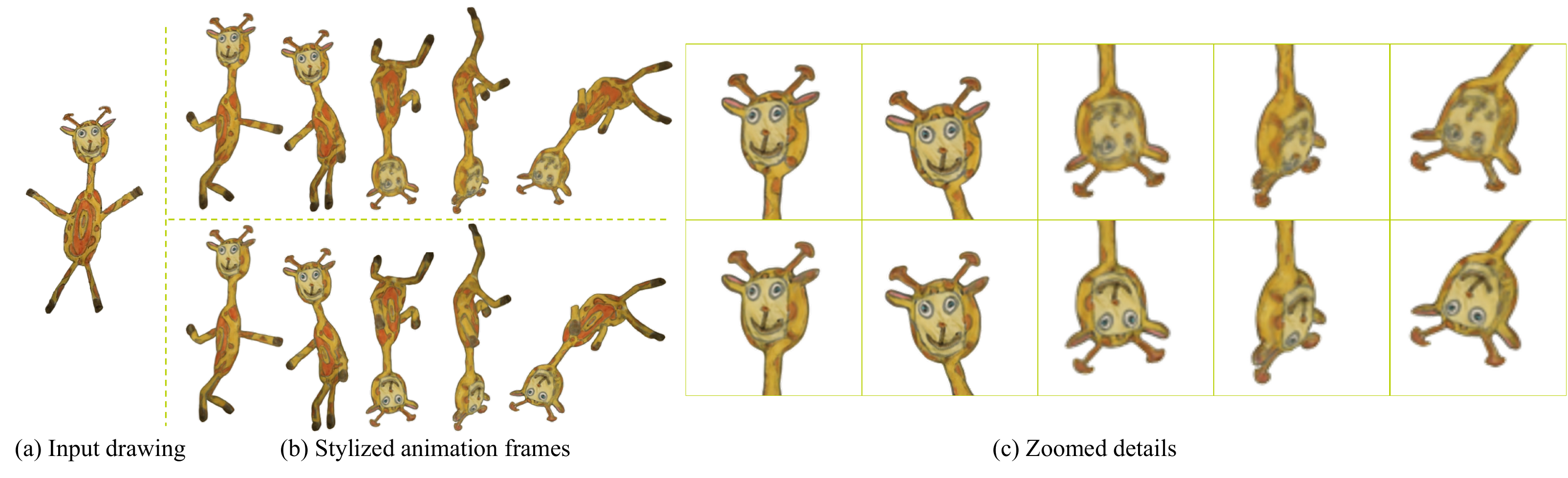}
  \vspace{-2mm}
  \caption{Comparison between w/o (Top) and w/ (Bottom) RIC convolution.}
  \vspace{-2mm}
  \label{fig:ab3}
\end{figure*}
We compare the results without (Top) and with (Bottom) RIC convolution in Fig.~\ref{fig:ab3}. Vanilla convolution cannot work well with the head-down animation frames since it lacks the ability to align the feature maps of a transformed image with those of its original. In contrast, RIC convolution alleviates this issue, enhancing the stability of results.

\section{Conclusion}
This paper has presented the first system \sysName~to generate vivid 3D animations by applying 3D motions to a single character drawing, while maintaining the contour style consistent with the input drawing. To reconstruct a 3D character as a proxy from the single drawing, our system borrows the reconstruction prior from a pre-trained image-to-3D diffusion model and makes it compatible with human-drawn drawings in terms of appearance and geometry.
For appearance improvement, we adopt a removal-then-restoration strategy to first remove the view-dependent contour lines and then render them back after retargeting the reconstructed character.
For shape refinement, we develop a cutting-and-thinning method to refine the slim structures represented by the single-line contours.

\begin{figure}[htbp]
  \centering
  \includegraphics[width=0.48\textwidth]{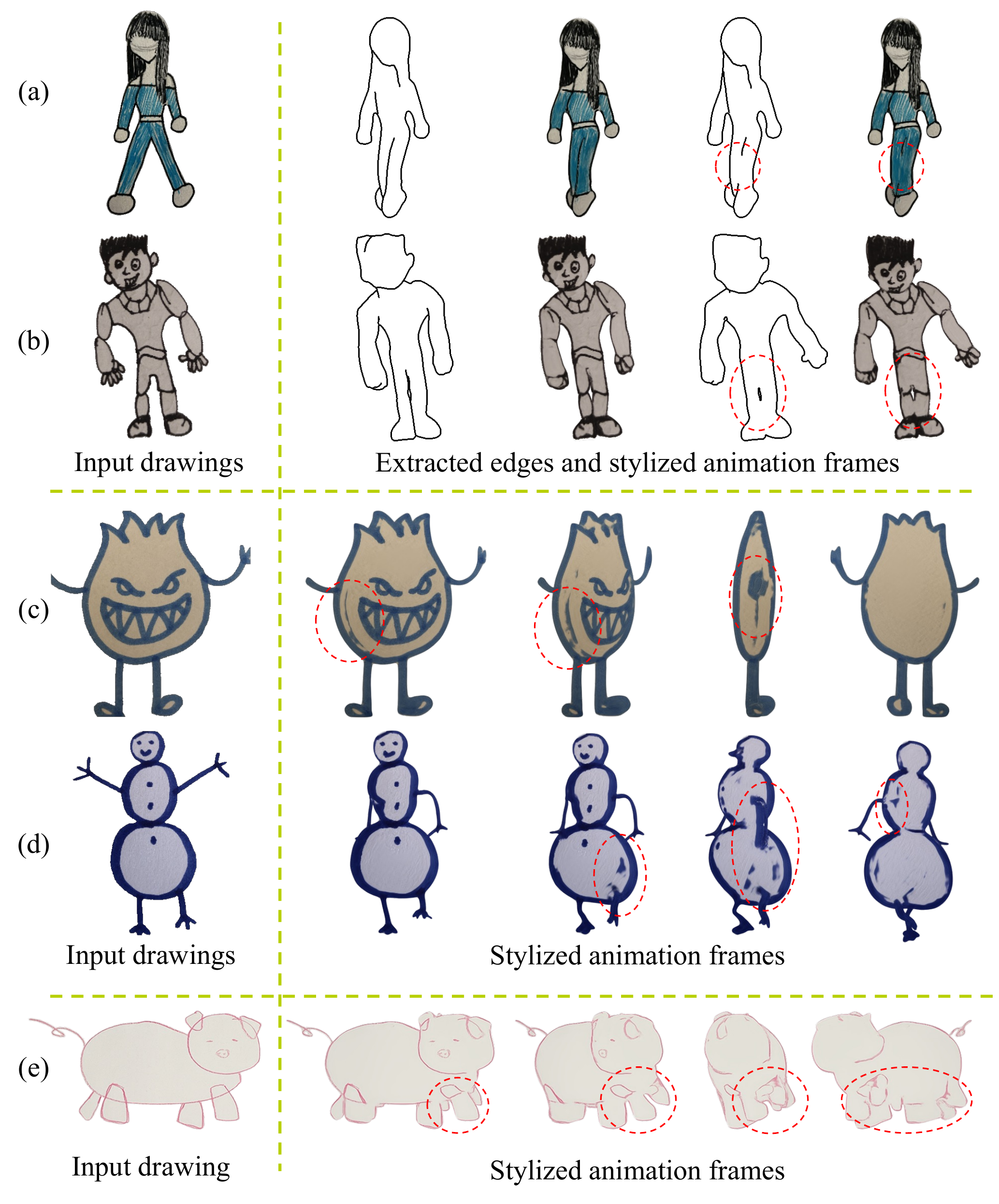}
  \vspace{-6mm}
  \caption{Some failure cases.}
  \vspace{-1mm}
  \label{fig:failure}
\end{figure}
While our system can produce visually plausible 3D animations, it still has several limitations.
First, we assume the characters in input drawings are approximately in a frontal A/T pose, without any self-occlusion. Some self-occlusion cases, such as crossed arms, would result in the reconstructed geometry exhibiting surface adhesion or merging of body parts.
Second, our contour rendering might produce artifacts when we extract edges with inappropriate thresholds (see Fig.~\ref{fig:failure} (a)-(b)). 
Third, when the contour lines of the input drawing are too thick, artifacts may appear in the generated results (see Fig.~\ref{fig:failure} (c)-(d)).
Additionally, it might produce artifacts when the drawings are highly abstract and far away from bipedal characters (see Fig.~\ref{fig:failure} (e)). 

In the future, we plan to extend our system for real-time performance, involving developing fast 3D reconstruction and auto-rigging methods, and investigate a generalized stylization network for acceleration. Although our method is not real-time, it benefits the community of character animation and non-photorealistic rendering. We envision its potential for drawing-to-animation storytelling by integrating text prompts and prototyping animation designs in cartoon films, games, VR/AR scenes, etc. We are also interested in extending our method to work for general human drawings, e.g., quadruped animals.

\begin{acks}
We thank the anonymous reviewers for their constructive comments and suggestions, which greatly enhanced the quality of this work. Additionally, we appreciate the participants who took the time to engage in the user study. This research was supported by a start-up fund by the Hong Kong University of Science and Technology (Project No. R9913) and the Chow Sang Sang Group Research Fund (Project No. 9229119).
\end{acks}

\bibliographystyle{ACM-Reference-Format}
\bibliography{files/reference}

\end{document}